\documentclass[aps,showpacs,preprintnumbers,amsmath, amssymb]{revtex4}

\usepackage{float}
\usepackage{graphics,epsfig}
\usepackage{graphicx}
\usepackage{dcolumn}
\usepackage{bm}

\begin{document}
\baselineskip=0.8 cm

\title{{\bf Large regular reflecting stars have no scalar field hair }}
\author{Yan Peng$^{1}$\footnote{yanpengphy@163.com}}
\affiliation{\\$^{1}$ School of Mathematical Sciences, Qufu Normal University, Qufu, Shandong 273165, China}

\vspace*{0.2cm}
\begin{abstract}
\baselineskip=0.6 cm
\begin{center}
{\bf Abstract}
\end{center}

We investigate the gravity system constructed with static
scalar fields coupled to asymptotically flat regular reflecting stars.
We consider the matter field's backreaction on the reflecting star.
We analytically show that there is an upper bound on the radius
of the reflecting star. When the star radius is above the bound,
the reflecting star cannot support the existence of scalar
field hairs. That means large reflecting stars
cannot have scalar field hairs.

\end{abstract}

\maketitle
\newpage
\vspace*{0.2cm}

\section{Introduction}

According to the no hair theorem \cite{Bekenstein,Chase,C. Teitelboim,Ruffini-1},
the asymptotically flat black hole
can be determined by the three conserved charges (the mass, angular momentum and charge of the black hole),
see references \cite{Hod-1}-\cite{Brihaye} and reviews \cite{Bekenstein-1,CAR}.
The belief on the no hair theorem was partly based on the physical argument that
exterior matter and radiation fields would either go away to the infinity or
be swallowed by the black hole horizon.
So no field can exist outside a black hole horizon, except when it is associated
with the three conserved charges of black hole spacetimes.

Whether there is also no hair theorem in the horizonless spacetime
is an interesting question to be answered.
Recently, it was found that the no scalar hair behavior
appears in the background of regular neutral reflecting stars.
It was firstly proved in \cite{Hod-6} that the
static scalar field cannot exist outside asymptotically flat
neutral compact reflecting stars without a horizon.
When considering the nonminimal coupling between the
massless scalar field and the gravity,
it was shown that the no scalar hair theorem
holds in certain range of the coupling model parameter \cite{Hod-7}.
And in the background of asymptotically dS gravity,
it was found that the massive scalar, vector and tensor field
cannot condense outside the regular neutral reflecting stars \cite{Bhattacharjee}.

Can the no hair theorem obtained in the neural horizonless gravity
still holds in the charged horizonless spacetime?
In the charged horizonless reflecting shell spacetime,
it was analytically found that the static scalar field cannot exist outside the shell
when the shell radius is large enough \cite{Hod-8,Hod-9,Yan Peng-1}.
Moreover, it was shown that charged horizonless reflecting stars
cannot support the existence of the scalar field when the star radius is above
an upper bound \cite{Hod-10,Yan Peng-2,Yan Peng-3,Yan Peng-4,Yan Peng-5}.
And the no scalar hair behavior also appears in
the large size regular star with other surface boundary conditions \cite{Yan Peng-6,Yan Peng-7}.
However, all the front discussion has been carried out in the probe limit.
In this work, we try to generalize the discussion by
considering the matter fields' backreaction on the metric and
also examine whether there are upper
bounds on hairy star radii.

This paper is structured as follows.
In section II, we construct the gravity system composed of a static scalar field
and a asymptotically flat regular reflecting star away from the probe limit.
In section III, we analytically show that there is an upper bound for
the reflecting star radius. Above the bound, the scalar field
cannot condense outside the star surface. Our summary is in the last section.

\section{The gravity model composed of scalar fields and reflecting stars}

We choose to study static asymptotically flat regular reflecting star spacetimes.
In Schwarzschild coordinates, the line element of the
spherically symmetric star is \cite{mr1,mr2}
\begin{eqnarray}\label{AdSBH}
ds^{2}&=&-g(r)e^{-\chi(r)}dt^{2}+g^{-1}dr^{2}+r^{2}(d\varphi^2+sin^{2}\varphi d\phi^{2}).
\end{eqnarray}
The solutions $\chi(r)$ and $g(r)=1-\frac{2m(r)}{r}$
only depend on the radial coordinate r.
Since we study the asymptotically flat spacetime,
there is $\chi(r)\rightarrow 0$ and $m(r)\rightarrow M$
as $r\rightarrow \infty$, where M is the total mass of the spacetime.

And the simple Lagrange density with scalar fields coupled to the charged star reads \cite{e5,ee,e6}
\begin{eqnarray}\label{lagrange-1}
\mathcal{L}=R-F^{MN}F_{MN}-|\nabla_{\alpha} \psi-q A_{\alpha}\psi|^{2}-\mu^{2}\psi^{2}.
\end{eqnarray}
Here $\psi(r)$ is the scalar field and $A_{\alpha}$ corresponds to the Maxwell field.
We define R, q and $\mu$ as Ricci curvature, scalar field charge and scalar field mass respectively.

We assume that the Maxwell field has only
the nonzero $t$ component in the form $A_{t}=\phi(r)dt$.
Then the equation of the Maxwell field is \cite{e5,ee,e6,e1,e2,e3,e4}
\begin{eqnarray}\label{BHg}
\phi''+(\frac{2}{r}+\frac{\chi'}{2})\phi'-\frac{q^2\psi^{2}}{2g}\phi=0.
\end{eqnarray}
At the infinity, the asymptotical behavior of the electric potential is $\phi=-\frac{Q}{r}$,
where Q is the total charge within the radius r \cite{Hod-10,Yan Peng-2}.
The scalar field equation is \cite{bl0,bl1,bl2,bl3,bl4,CW}
\begin{eqnarray}\label{BHg}
\psi''+(\frac{2}{r}-\frac{\chi'}{2}+\frac{g'}{g})\psi'+(\frac{q^2e^{\chi}\phi^2}{g^2}-\frac{\mu^2}{g})\psi=0.
\end{eqnarray}
And the metric equations are \cite{e5,ee,e6}
\begin{eqnarray}\label{BHg}
\chi'+r\psi'^2+\frac{q^2re^{\chi}\phi^2\psi^2}{g^2}=0,
\end{eqnarray}
\begin{eqnarray}\label{BHg}
g'+\frac{g}{r}-\frac{1}{r}+rg[\frac{1}{2}\psi'^2+\frac{e^{\chi}\phi'^2}{g}+\frac{q^2e^{\chi}\phi^2\psi^2}{2g^2}+\frac{\mu^2\psi^2}{2g}]=0.
\end{eqnarray}

In the linear case, we can simply set $\psi(r)=0$ in Eq.(3), Eq.(5) and Eq.(6).
The metric solutions are $g(r)=1-\frac{2M}{r}+\frac{Q^2}{r^2}$, $\chi=0$
and the Maxwell field outside the star is $\phi(r)=-\frac{Q}{r}$,
where $Q$ is the star charge and $M$ is the star mass.
We define $r_{s}$ as the star radius.
In the limit case of $Q,M\ll r_{s}$ while $qQ$ fixed,
non-trivial solutions of Eq.(4) were analytically obtained in \cite{Hod-8,Hod-9,Yan Peng-1}.
The relations $Q,M\ll r_{s}$ and $g(r)=1-\frac{2M}{r}+\frac{Q^2}{r^2}$
mean that exterior regions outside the
star is assumed to be flat. We should point out that
the presence of a charge $Q$ is crucial for the
existence of the analytical solution in terms of
Bessel functions. With the nonzero term $qQ$, $Q$ appears in the
scalar field equation \cite{Hod-8,Hod-9,Yan Peng-1}.
As a further step, for nonzero values $q$, $Q$ and $M$,
we numerically obtained non-trivial
solutions of Eq.(4) in the linear limit \cite{Yan Peng-2}.

However, all front discussions have been carried out without scalar fields' backreaction.
When considering scalar fields' backreaction
on the metric, Eq.(4) is coupled with Eq.(3), Eq.(5) and Eq.(6).
In this case of nonlinear coupled equations, approaches in \cite{Hod-8,Hod-9,Yan Peng-1,Yan Peng-2} fail
and more precise numerical methods are needed.
In this work, we analytically show that the non-trivial scalar field solution
of nonlinear equations cannot exist when the star radius is above an upper bound.
In other words, we prove no scalar field hair theorem
for large reflecting stars in the nonlinear regime.

At the star surface, we take the scalar reflecting condition that
the scalar field vanishes. Around the infinity,
the scalar field asymptotically behaves in the form $\psi\sim A\cdot\frac{1}{r}e^{-\mu r}+B\cdot\frac{1}{r}e^{\mu r}$,
with A and B as integral constants.
In order to obtain the physical solution, we set $B=0$.
And the scalar field satisfies boundary conditions
\begin{eqnarray}\label{InfBH}
&&\psi(r_{s})=0,~~~~~~~~~\psi(\infty)=0.
\end{eqnarray}

The mass $m(r)$ within the radius r is given by
\begin{eqnarray}\label{AdSBH}
m(r)=\int_{0}^{r}4\pi r'^{2}\rho(r')dr',
\end{eqnarray}
where $\rho=-T^{t}_{t}$ is the energy density.
For the probe charged star, there is $m(r)=M-\frac{Q^2}{2r}$.
In this work, we are interested in general asymptotically flat star with backreaction of matter fields.
Considering the facts that the scalar field asymptotically goes to zero as $\psi\sim\frac{1}{r}e^{-\mu r}$
and charged Maxwell fields are associated with the $-\frac{Q}{r}$ asymptotic behavior,
we deduce that the Maxwell field dominates the energy density and $\rho=-T^{t}_{t}\thicksim \frac{1}{r^4}$
around the infinity \cite{mr1,bl5}. So there is
\begin{eqnarray}\label{AdSBH}
m'(r)=4\pi r^{2}\rho(r)\rightarrow 0
\end{eqnarray}
as $r\rightarrow\infty$.

\section{Upper bounds on radii of scalar hairy reflecting stars}

With a new function $\tilde{\psi}=\sqrt{r}\psi$,
the equation (4) can be expressed as
\begin{eqnarray}\label{BHg}
r^2\tilde{\psi}''+(r-\frac{r^2\chi'}{2}+\frac{r^2g'}{g})\tilde{\psi}'+(-\frac{1}{4}-\frac{rg'}{2g}+\frac{r^2q^2e^{\chi}\phi^2}{g^2}-\frac{\mu^2r^2}{g})\tilde{\psi}=0.
\end{eqnarray}

According to (7), boundary conditions of $\tilde{\psi}$ are
\begin{eqnarray}\label{InfBH}
&&\tilde{\psi}(r_{s})=0,~~~~~~~~~\tilde{\psi}(\infty)=0.
\end{eqnarray}
Then the function $\tilde{\psi}$ must possess at least
one extremum point $r=r_{peak}$ above the star surface $r_{s}$.
At this extremum point, there is the following characteristic relation
\begin{eqnarray}\label{InfBH}
\{ \tilde{\psi}'=0~~~~and~~~~\tilde{\psi} \tilde{\psi}''\leqslant0\}~~~~for~~~~r=r_{peak}.
\end{eqnarray}

With relations (10) and (12), we arrive at the following inequality
\begin{eqnarray}\label{BHg}
-\frac{1}{4}-\frac{rg'}{2g}+\frac{r^2q^2e^{\chi}\phi^2}{g^2}-\frac{\mu^2r^2}{g}\geqslant0~~~for~~~r=r_{peak}.
\end{eqnarray}

It can be transformed into
\begin{eqnarray}\label{BHg}
\mu^2r^2g\leqslant r^2e^{\chi}q^2\phi^2-\frac{rgg'}{2}-\frac{1}{4}g^2~~~for~~~r=r_{peak}.
\end{eqnarray}

At the infinity, the asymptotical behaviors of the solutions are $\phi\rightarrow -\frac{Q}{r}$
and $e^{\chi}\rightarrow 1$.
So a constant $R_{1}$ exists and for any $r\geqslant r_{s}\geqslant R_{1}$,
the relations $-\frac{2Q}{r}<\phi<-\frac{Q}{2r}$ and $e^{\chi}\leqslant \frac{5}{4}$ hold.
So there is
\begin{eqnarray}\label{BHg}
\mu^2r^2g\leqslant 5q^2Q^2-\frac{rgg'}{2}-\frac{1}{4}g^2~~~for~~~r=r_{peak}.
\end{eqnarray}

According to (9), we have the relation
\begin{equation}\label{BHg}
\begin{split}
rgg'=r(1-\frac{2m(r)}{r})(1-\frac{2m(r)}{r})'=r(1-\frac{2m(r)}{r})(\frac{2m(r)}{r^2}-\frac{2m'(r)}{r})\\
=(1-\frac{2m(r)}{r})(\frac{2m(r)}{r}-2m'(r))
\rightarrow (1-\frac{2M}{r})(\frac{2M}{r}-2m'(r))\rightarrow 0
\end{split}
\end{equation}
as $r\rightarrow\infty$. Then there is a constant $R_{2}$ and for any $r\geqslant r_{s}\geqslant R_{2}$,
the inequality $-2q^2Q^2<rgg'<2q^2Q^2$ holds.
So there is the relation
\begin{eqnarray}\label{BHg}
\mu^2r^2g\leqslant 6q^2Q^2-\frac{1}{4}g^2\leqslant 6q^2Q^2~~~for~~~r=r_{peak}
\end{eqnarray}
with $r_{s}\geqslant R_{1}$ and $r_{s}\geqslant R_{2}$.

Since $g=1-\frac{2m(r)}{r}\rightarrow 1-\frac{2M}{r} \rightarrow 1$ as $r\rightarrow\infty$,
there is a constant $R_{3}$ and for any $r\geqslant r_{s}\geqslant R_{3}$,
there is the relation $g>\frac{1}{2}$. So we have

\begin{eqnarray}\label{BHg}
\frac{\mu^2r_{s}^2}{2}\leqslant \frac{\mu^2r^2}{2}\leqslant \mu^2r^2g\leqslant 6q^2Q^2~~~for~~~r=r_{peak}
\end{eqnarray}
with $r_{s}\geqslant R_{1}$, $r_{s}\geqslant R_{2}$ and $r_{s}\geqslant R_{3}$.

We assume that star radii satisfy $r_{s}\geqslant R_{1}$, $r_{s}\geqslant R_{2}$ and $r_{s}\geqslant R_{3}$,
otherwise we have an upper bound
\begin{equation}
\mu r_{s} \leqslant max \left\{\mu R_{1},\mu R_{2},\mu R_{3} \right\}.
\end{equation}

According to (18) and $r_{peak}\geqslant r_{s}$, there is
\begin{eqnarray}\label{BHg}
\mu r_{s}\leqslant \mu r_{peak}\leqslant 2\sqrt{3}qQ.
\end{eqnarray}

Our analysis shows that the hairy star radius is below the bounds (19) or (20).
That is to say the hairy star radii can be divided into two cases

~~~~case 1:~~~$\mu r_{s} \leqslant  max \left\{\mu R_{1},\mu R_{2},\mu R_{3} \right\}$;

~~~~case 2:~~~$\mu r_{s} \geqslant  max \left\{\mu R_{1},\mu R_{2},\mu R_{3} \right\}$ with $\mu r_{s}\leqslant  2\sqrt{3}qQ$.

In all, we obtain upper bounds for hairy star radii as
\begin{equation}
\mu r_{s} \leqslant max \left\{2\sqrt{3}qQ,\mu R_{1},\mu R_{2},\mu R_{3} \right\},
\end{equation}
with dimensionless quantities according to the symmetry of equations (3-6)
\begin{eqnarray}\label{BHg}
r\rightarrow k r~,~~~~~~~~~~ \mu\rightarrow \mu/k~,~~~~~~~~~~ q\rightarrow q/k~.
\end{eqnarray}

When the star radius is above the bound (21), the static scalar field cannot condense outside
the static asymptotically flat spherically symmetric regular reflecting star.
That means big reflecting star cannot support
the existence of massive scalar field hair.
In fact, there are also similar properties in black hole gravities.
According to the no short hair conjecture, the hairy black hole horizon
is below the upper bound $r_{H}<\frac{2}{3}(\eta)^{-1}$
with $\eta$ as the exterior field mass \cite{ub1}.
In fact, the numerical results also suggest
that big black holes tend to have no massive hair \cite{ub2} .

\section{Conclusions}

We studied static massive scalar field condensations outside
static asymptotically flat spherically symmetric regular reflecting stars.
We constructed a complete gravity model by considering
the matter field's backreaction on the background.
We provided upper bounds on the star radius in the form
$\mu r_{s} \leqslant max \left\{2\sqrt{3}qQ,\mu R_{1},\mu R_{2},\mu R_{3} \right\}$,
where $\mu$ is the scalar field mass, q is the charge coupling parameter, Q is
the total charge and $R_{i}$ depends on the gravity theories.
When the star radius is above the bound, the static massive
scalar field cannot condense outside the reflecting star.
That means the large regular reflecting star
cannot have massive scalar field hairs
in the asymptotically flat gravity.

\begin{acknowledgments}

We would like to thank the anonymous referee for the constructive suggestions to improve the manuscript.
This work was supported by the Shandong Provincial Natural Science Foundation of China under Grant
No. ZR2018QA008.

\end{acknowledgments}

\end{document}